\documentclass[amssymb,superscriptaddress,twocolumn,prd]{revtex4}
\usepackage{graphicx}
\def \be{\begin{equation}}
\def \ee{\end{equation}}
\def \bea{\begin{eqnarray}}
\def \eea{\end{eqnarray}}

\newcommand{\beq}[1]{\begin{eqnarray}\label{#1}}
\newcommand{\eeq}{\end{eqnarray}}

\begin{document}
 \pagestyle{plain}

 \title{A Question About \\
 Standard Cosmology and Extremely Dense Stars' Collapsing}

\author{Ding-fang Zeng and Yi-hong Gao}
\email{dfzeng@itp.ac.cn} \email{gaoyh@itp.ac.cn}
\affiliation{Institute of Theoretical Physics, Chinese Academy of
Science.}
 \begin{abstract}
 We ask if the conventional variable separation techniques in
 the studying of standard cosmology and the collapsing of
 extremely dense stars introduce Newton's absolute space-time
 concepts. If this is the case, then a completely relative
 cosmology is needed. We build the basic frame-works for such a
 cosmology and illustrate that, the observed luminosity-distance
 v.s. red-shift relations of supernovaes can be explained naturally
 even without any conception of dark energies.
 \end{abstract}


 \maketitle

 \section{Our Questions}

 Either in the standard cosmology, or in the
 a spherical collapsing star, we are told to start with
 a homogeneous and isotropic fluid ball,
 write down the dynamic equation describing its
 evolutions as,
 \beq{}
 G_{\mu\nu}=-8\pi GT_{\mu\nu},\label{EinsteinEquation}
 \eeq
 and the general expressions for energy momentum tensor,
 $T_{\mu\nu}=\rho
 u_{\mu}u_{\nu}+p(g_{\mu\nu}+u_{\mu}u_{\nu})$, in the co-moving
 reference frame is written as
 \beq{}
 &&\hspace{-3mm}T_{\mu\nu}=\textrm{dial}(\rho,p,p,p)\label{SCosmologyEMT}\nonumber\\
 &&\hspace{7mm}\textrm{if no radiation and/or}\nonumber\\
 &&\hspace{7mm}\textrm{ no dark energy is involved}\nonumber\\
 &&\hspace{4mm}=\textrm{dial}(\rho,0,0,0)
 \eeq
 In solving eq(\ref{EinsteinEquation}), we are told to start from
 the metric ansaltz
 \beq{}
 ds^2=-dt^2+U(t,r)dr^2+V(t,r)(d\theta^2+\textrm{sin}^2\theta
 d\phi^2),
 \label{MetricAnsaltz1}
 \eeq
 using variable separation techniques, setting
 \beq{}
 &&U(t,r)=a^2(t)f(r),\nonumber\\
 &&V(t,r)=a^2(t)\cdot r^2,\nonumber\\
 &&f(r)=\frac{1}{1-kr^2},k=0,\pm1
 \label{FactorizeUVfunc}
 \eeq
 and finally obtain Friedmann Equation
 \beq{}
 (\frac{\dot{a}}{a})^2+\frac{k}{a^2}=\frac{8\pi G}{3}\rho
 \eeq

 Our question is, can the factorized ($U$ and $V^\prime$s dependence on
 $t$ and $r$) metric
 eq(\ref{MetricAnsaltz1})+(\ref{FactorizeUVfunc})
 be used to describe our observable universe or
 the inner space-time of a collapsing star?
 The logics in this metrics
 is, we have a mathematically well defined co-moving reference
 mesh, on which all matter particles can be put instantaneously
 and independently, and at the putting epoch, all the matter
 particles are at rest and of no inter-gravitation at all. Since
 otherwise, the expansion or contraction of basic particles in the
 system will produce pressures.

 The $U(t,r)$ and $V(t,r)$ factorizable metric
 eqs(\ref{MetricAnsaltz1})+(\ref{FactorizeUVfunc}) is only a
 special kind of
 solutions of Einstein equation. This solution requires us to have
 a globally defined scale factor $a(t)$ and a co-moving reference mesh in the
 universe or in the collapsing star. At the initial time of
 ``universe creation'', such
 as inflation beginning, Ekpyrotic universe's
 Brane-Brane-collision-point, or the collapsing beginning epoch
 of the extremely dense stars, all the objects in
 the system were put
 on the co-moving mesh instantaneously and
 independently, but homogeneously
 and isotropically. For this reason, all these objects are
 of no inter-gravitation, and no pressures will be introduced at the initial time.
 On the contrary, the $U(t,r)$ and $V(t,r)$ non-factorizable
 solutions also exist and in those solutions,
 all the objects in the system are all
 inter-gravitated from the the beginning. We have no
 epochs at which those objects's inter-gravitation can be set to
 zero, hence no epochs at which the pressures originated from
 these objects' expanding or falling down can be neglected.

 Phylosophically speaking, assuming the metric function of
 the system can be factorized as $U(t,r)=a^2(t)f(r)$, probably
 introduces Newton's absolute space-time in our studys.
 In this space-time concepts,
 all the objects in the system were put on the
 co-moving meshes (absolute space) instantaneously and
 independently at the initial epochs. In the following
 evolutions, the distances between any two space points
 increase or decreases at the same speed, because the
 scale factor $a(t)$ (absolute time) is defined
 globally in the system. So if we were put on a given point in the
 system, we must see some points are running away from us at
 infinitely speed, as long as those points are infinitely away from
 us. Obviously, this contradicts special relativity.

 \section{Maximumly Symmetric Space?}

 Some people may tell us the fact that our universe should be described by
 Freidmann-Robertson-Walker metric is the requirement of symmetry.
 It has nothing to do with other things such as those we stated in the
 previous section. To this argument, we would like to point out
 that, the symmetries referred to by these peoples are symmetries
 of an absolute space, its operating definition depends on the existance
 of a signal which can be infinitely speedy. So,
 our question may be asked as, is
 cosmological principle in the standard cosmology expressed
 in an anti-relativity way? (The case of extremely dense stars' collpasing
 theory has similar problem.) Our meaning is, if special
 relativity is considered in the definition of homogeneity
 and isotropy, then we must consider the finiteness of signal
 transferring speed, i.e., the homogeneity and isotropy must be a
 statement which can be tested by signals no more speedy
 than light.

 So, special relativity may suggest that, to build a completely
 relative cosmology, we may have to find a new expression of
 cosmological principles which is consistent with the finiteness of
 light's tranferring speed.

 \section{Constructure A Completely
 Relative Theory of Cosmology and Black Holes' Formation}

 If we discard the usual definition of cosmological principle or
 similar expressions in the extremely dense stars collapsing case
 which is based on the existence of
 signals which have infinitely large tranferring speeds, we will
 have no co-moving mesh anymore on which the basic particles in our systems
 can be put instantaneously and independently, and the physical
 distances between any two matter particles in our systems can not be
 factorized as the multiplying of an only-time-dependent scale factor and an
 absolutely defined co-moving distance. In this case the function
 $U(t,r)$ and $V(t,r)$ is non-factorizable as in
 eq(\ref{FactorizeUVfunc}).

 Techniquely analyzing, the
 Hubble recession of the basic molecules of cosmological gases or
 the falling down of the matter particles in the extremely dense
 stars make the energy momentum tensor of the system
 non-diagonal, hence make the
 function $U(t,r)$ and $V(t,r)$ non-factorizable.
 To emphasize this point,
 let us rewrite the metric of our systems in the following,
 \beq{}
 ds^2=-dt^2+U(t,r)dr^2+V(t,r)(d\theta^2+\textrm{sin}^2\theta
 d\phi^2).\label{MetricAnsaltz2}
 \eeq
 Denote the four velocity of basic molecules of the cosmological
 gases or that of the matter particles in the extremely dense stars as
 \beq{}
 &&\hspace{-3mm}u^{\mu}=(u^0,u^1,0,0),\ \textrm{but}\nonumber\\
 &&\hspace{5mm}\frac{u^1}{u^0}=v,\nonumber\\
 &&\hspace{5mm}-(u^0)^2+(u^1)^2U(t,r)=-1,\label{FourVelocityDefinition}
 \eeq
 where $v$ is appropriate Hubble recession or falling down
 velocity of the basic particles in our studying system. It is an
 observable quantity and its value depends on $(t,r)$. The usual hubble parameter is defined
 as
 \beq{}
 H=\frac{1}{r}v=\frac{1}{r}\frac{dr}{dt}.
 \eeq
 From eqs(\ref{FourVelocityDefinition})
 we can solve
 \beq{}
 u^0=\frac{1}{\sqrt{1-Uv^2}},u^1=\frac{v}{\sqrt{1-Uv^2}}.
 \label{FourVelocityExpression}
 \eeq
 So the energy momentum tensor describing our cosmological
 gas or the matter particles in the extremely dense stars is
 \beq{}
 &&\hspace{-3mm}T^{\mu\nu}=\rho u^\mu u^\nu+p(g^{\mu\nu}+u^\mu u^\nu),\nonumber\\
 &&\hspace{7mm}
 [p=\frac{1}{3}\rho v^2, \textrm{we are not sure if}\nonumber\\
 &&\hspace{9mm}\textrm{$\rho$ should depend on $(t,r)$ or not}\nonumber\\
 &&\hspace{0mm}=\rho
 \left[
 \begin{array}{cccc}
 \frac{1+\frac{Uv^4}{3}}{1-Uv^2}&\frac{v(1+\frac{v^2}{3})}{1-Uv^2}&0&0\\
 \frac{v(1+\frac{v^2}{3})}{1-Uv^2}&\frac{(1+3U)v^2}{3U(1-Uv^2)}&0&0\\
 0&0&\frac{v^2}{3V}&0\\
 0&0&0&\frac{v^2}{3V\textrm{sin}^2\theta}
 \end{array}
 \right].\label{NewEMT}
 \eeq
 In the standard cosmology, an $r$ dependent matter
 density violates cosmological principle, which says that
 the matter density in our universe is homogeneous and
 isotropic on large scales. While in the conventional
 black hole theory, a $(t,r)$ dependent matter distribution
 inside the horizon of the black holes violates ``no-hair''
 theorem.

 However, we are intending to believe that the energy
 density $\rho$ appearing
 in the above energy momentum tensor should depend on
 $t$ and $r$.
 Because the standard cosmology definition of
 homogeneity and isotropy (or similar experssions
 in the extremely collapsing stars have the same problem) depends on
 a globally defined co-moving reference mesh. It is on this
 co-moving reference mesh that the matter distribution in our universe
 are homogeneous and isotropic. We have explained this globally defined
 co-moving reference mesh may just be Newton's absolute space, so the
 homogeneity and isotropy definitions on
 this reference mesh cannot
 be tested by experiment. The definition which can be tested by
 experiment should depends on an experiment tool, such as light.
 So we have to introduce a four velocity
 \beq{}
 &&\hspace{-3mm}c^\mu=(c^0,c^1,0,0),\ \textrm{but}\nonumber\\
 &&\hspace{5mm}\frac{c^1}{c^0}=1,\nonumber\\
 &&\hspace{5mm}-(c^0)^2+(c^1)^2U(t,r)=0,\label{FourVelocityLight}
 \eeq
 Only observed by $c^\mu$, our universe is homogeneous and
 isotropic, and the black hole are ``no-hair'', i.e.,
 \beq{}
 &&\hspace{-3mm}T^{\mu\nu}c_{\mu}c_{\nu}=\rho_{aver},\nonumber\\
 &&\hspace{-3mm}T^{\mu\nu}(g_{\mu\nu}+c_{\mu}c_{\nu})=0,\nonumber\\
 &&\textrm{ $\rho_{aver}$ only
 depends}\,\nonumber\\
 &&\hspace{12mm}\textrm{on $t$, or neither $t$ nor $r$.}
 \label{NewCosmoPrinciple}
 \eeq
 Eq(\ref{NewCosmoPrinciple}) is our definition of cosmological
 principle or similar statement about the matter distributions in
 the extremely collapsing stars.

 It is worth noting that superficially looking
 eq(\ref{FourVelocityLight})
 means that $U(t,r)==1$ (we will use ``$==$'' denoting identical
 relations and ``$\equiv$'' denoting definition relations).
 Is this the case? No, this only means that
 \beq{}
 U(t,r)_{r=t}=1,\label{LightSpeedInvariance}
 \eeq
 i.e., along the trace of light
 $c^1=c^0$ hence $dr=dt$ and hence $r=t$, the function $U(t,r)$ takes
 value $1$. In some sense, this can be looked as an assumption
 which has counterpart in standard cosmology as,
 $U(t,r)^\prime$s dependence on $t$ and $r$ is factorzed globally.

 Starting from eq(\ref{NewCosmoPrinciple}), explicit calculations will give us
 \beq{}
 &&\hspace{-3mm}\rho_{aver}(t)=\rho(t,r)(1+\frac{v^2}{3})(u\cdot c)^2,\nonumber\\
 &&\hspace{-3mm}0=\rho(t,r)[(-1+v^2)+(1+\frac{v^2}{3})(u\cdot
 c)^2],\label{NewCosmoPrincipleExplicit}
 \eeq
 From which we can solve
 \beq{}
 &&\hspace{-3mm}\rho(t,r)=\frac{\rho_{aver}(t)}{(1+\frac{v^2}{3})(u\cdot
 c)^2},\nonumber\\
 &&\hspace{-3mm}v(t,r)=\frac{2U+F-(9U^4-18U^3-19U^2-36U)F^{-1}}{3(U^2-3U)},
 \nonumber\\
 &&\hspace{-3mm}F=\left[\frac{1}{2}(
 216U^2+1564U^3-864U^4+108U^5\right.\nonumber\\
 &&\hspace{8mm}\left.
 +\sqrt{4(-36U-19U^2-18U^3+9U^4)^3}\right.\nonumber\\
 &&\hspace{12mm}
    \left.\overline{+(216U^2+1564U^3-864U^4+108U^5)^2}
 )
 \right]^{\frac{1}{3}}.
 \nonumber\\
 \label{NewCosmoPrincipleDerivation}
 \eeq
 Let us make a little plume of our logics here. In
 eq(\ref{FourVelocityExpression}), we express the four velocity of
 the basic particles in our systems (universe or the inside space of
 extremely dense stars) in terms of $U(t,r)$ and $v(t,r)$, in
 eq(\ref{NewEMT}) we express the energy momentum tensor
 describing the systems in terms
 of $U(t,r)$ and $v(t,r)$, while in
 eqs(\ref{NewCosmoPrinciple})+(\ref{NewCosmoPrincipleExplicit})+(\ref{NewCosmoPrincipleDerivation})
 according to cosmological principle (or similar statement such as
 black holes' ``no-hair'' theorem), we derive out the explicit dependence of
 $v(t,r)$ on $U(t,r)$. So if we substitute the results in
 eq(\ref{NewCosmoPrincipleDerivation}) back into eq(\ref{NewEMT}),
 we will have an energy momentum tensor expressed in terms of
 $U(t,r)$ and $\rho_{aver}(t)$, which can be solved from
 Einstein equations and energy momentum conservation law uniquely.

 \section{Energy Momentum Conservation And Einstein Equations}

 In standard cosmology or the conventional black hole
 formation thery, energy momentum conservation has simple
 forms, for the no-radiation and no-dark-energy cosmology, it is $d(\rho
 [a\cdot r_{co}]^3)=0$; for the with-radiation or/and with-dark-energy case, $d(\rho
 [a\cdot r_{co}]^3)+pd[a\cdot r_{co}]^3=0$, where $a$ is the
 only-time-dependent scale factor of the universe, while $r_{co}$
 is the time-independent co-moving distance.

 However, if what we analyzed in the previous sections is the
 fact, then our current standard cosmology must be some
 kinds of approximation which is applicable only in short time
 evolution or small space phenomenology of the strict relative
 cosmology we proposed in the above section. In this case, the
 energy momentum conservation law $T^\mu_{\nu;\mu}=0$ can be
 integrated to give
 \beq{}
 \int_0^r\frac{\rho_{aver}(t)(1-Uv^2)}{(1+\frac{v^2}{3})(1-Uv)^2}\sqrt{U(t,r)}V(t,r)dr=const.
 \label{intEMConserve}
 \eeq
 We do not consider radiation and dark energy in our new
 frame-work of cosmology in this paper.
 Einstein equaiton
 \beq{}
 R_{\mu\nu}=-8\pi G(T_{\mu\nu}-\frac{1}{2}g_{\mu\nu}T)
 \eeq
 has the following
 non-trivial component,
 \beq{}
 &&\hspace{-3mm}\left[
 R_{01}=\frac{\dot{V}^\prime}{V}-\frac{V^\prime
 \dot{V}}{2V^2}-\frac{\dot{U}V^\prime}{2UV}
 \right]=\nonumber\\
 &&\hspace{12mm}8\pi G\rho_{aver}(t)
 \frac{Uv}{(1-Uv)^2},
 \nonumber\\
 \nonumber\\
 \nonumber\\
 &&\hspace{-3mm}\left[
 R_{00}=\frac{\ddot{U}}{2U}+\frac{\ddot{V}}{V}
 -\frac{\dot{U}^2}{4U^2}-\frac{\dot{V}^2}{2V^2}\right]=
 \nonumber\\
 &&\hspace{5mm}
 8\pi G\rho_{aver}(t)\left[
 \frac{(1-\frac{v^2}{3})(1-Uv^2)}{2(1+\frac{v^2}{3})(1-Uv)^2}-\frac{1}{(1-Uv)^2}
 \right],
 \nonumber\\
 \nonumber\\
 \nonumber\\
 &&\hspace{-3mm}\left[
 R_{11}=\frac{V^{\prime\prime}}{V}-\frac{V^{\prime
 2}}{2V^2}-\frac{U^\prime
 V^\prime}{2UV}-\frac{\ddot{U}}{2}+\frac{\dot{U}^2}{4U}-\frac{\dot{U}\dot{V}}{2V}
 \right]=\nonumber\\
 &&\hspace{5mm}8\pi G\rho_{aver}(t)\left[
 \frac{U(1+v^2)(1-Uv^2)}{2(1+\frac{v^2}{3})(1-Uv)^2}-\frac{U}{(1-Uv)^2}
 \right],
 \nonumber\\
 \nonumber\\
 \nonumber\\
 &&\hspace{-3mm}\left[
 R_{22}=-1+\frac{V^{\prime\prime}}{2U}-\frac{V^\prime
 U^\prime}{4U^2}-\frac{\ddot{V}}{2}-\frac{\dot{V}\dot{U}}{4U}
 \right]=\nonumber\\
 &&\hspace{10mm}-8\pi G\rho_{aver}(t)
 \left[
 \frac{V(1-\frac{v^2}{3})(1-Uv^2)}{2(1+\frac{v^2}{3})(1-Uv)^2}
 \right].
 \nonumber\\
 \label{EinsteinEquationComponents}
 \eeq

 So in our frame-work of cosmology,
 to predict experiments such as super-novaes'
 luminosity-distance v.s. red-shift relations,
 we need substitute the expressions of $v(t,r)$ in
 eq(\ref{NewCosmoPrincipleDerivation}) into
 eq(\ref{EinsteinEquationComponents}) and combine with
 eqs(\ref{LightSpeedInvariance})+(\ref{intEMConserve}),
 then solve the resulting equations to get the function $U(t,r)$.
 As long as $U(t,r)$ is obtained, we can get the Hubble recession
 velocity $v$ v.s. $(t,r)$ relation, from which the super-novaes'
 luminosity-distance v.s. red-shift relation should be calculated.

 Obviously, the forms of eq(\ref{intEMConserve})+(\ref{EinsteinEquationComponents})
 are so complicated that almost no attempts of directly solving
 the system is to be successful. We are now working by a different
 strategy, that is, guess a solution such as
 $ds^2=-dt^2+e^{\frac{t}{r}-1}(dr^2+r^2d\Omega_2^2)$, note
 this solution has satisfy
 eq(\ref{LightSpeedInvariance}) already, then
 using(\ref{EinsteinEquationComponents}) to get the function $\rho_{aver}(t)$,
 and check if the resulting function $\rho_{aver}(t)$ can satisfy
 eq(\ref{intEMConserve}) or not. Of course, if we
 require the solution come back to the usual
 Friedmann-Robertson-Walker metric, we can add extra requirements
 on our guess starting metric functions $U(t,r)$ and $V(t,r)$.
 We have not obtained definite
 conclusions on this aspect to provide here. But we think this is
 a challengeful and meaningful tasks both on physics and
 mathematics.

 \section{Is Our Universe Accelerately Expanding?}

 \begin{figure}
 \includegraphics[]{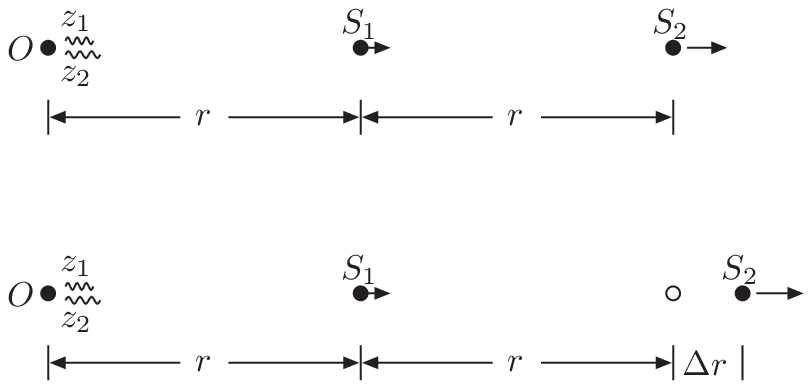}
 \vspace{5mm}
 \caption{
 Inferring that our universe is expanding accelerately:
 if $S_1$ and $S_2$ are two supernovaes $r$ and $2r$ away from
 us and if we observe photons from them have
 red-shift $z_1$ and $z_2$ (upper figure), then if we observe photons from two
 supernovaes also with red-shift $z_1$ and $z_2$,
 but are $r$ and $2r+\Delta r$ away from us
 (down figure, $\Delta r>0$), in standard cosmology we infer that our universe
 is accelerately expanding.
 }
 \label{accelerateExplanation}
 \end{figure}

 Inferring that our universe is accelerately expanding from
 super-novae's luminosity-distance v.s. red-shift relation is a
 well known and widely accepted conclusion, we illustrate the basic
 principles in FIG.\ref{accelerateExplanation}.

 In standard cosmology, if $S_1$ and $S_2$ are two supernovaes
 with distances $r$
 and $2r$ away from us, then if $S_1$ are recessing away from us
 at speed $v$, we infer that $S_2$ is doing so at speed $2v$,
 please see the upper figure of FIG.\ref{accelerateExplanation}.
 This logics of reasoning is based on the
 Friedmann-Robertson-Walker metrics, the physical distance
 $r=a(t)\cdot r_{co}$, $a(t)$ is defined globally.

 Our view point is, on large scales, we cannot write the metrics
 of our universe as Friedmann-Robertson-Walker type, we cannot
 define a globally meaningful scale factor $a(t)$ so that physical
 distance can be factorized as $r=a(t)\cdot r_{co}$. If $S_1$ is
 a supernovae with distance $r$ away from us and it is recessing from us at
 speed $v$, we can only infer that if $S_2$ is $2r$ away from us,
 $S_2$ is recessing away from $S_1$ at speed $v$, while the
 relative recessing speed between $S_2$ and us should be
 calculated according to the velocity addition law of special
 relativity, $v_{2O}=\frac{2v}{1+v^2}$. Obviously this is smaller
 than $2v$. To get a relative speed $v_{2O}=2v$, $S_2$ must be put on $2r+\Delta
 r$ away from us, please see the down figure of
 FIG.\ref{accelerateExplanation}.

 So if what we say (standard cosmology's space-time is
 Newton's absolute one) is correct, then the luminosity-distance v.s.
 red-shift relations detected in the current observations may do not
 mean that our universe is accelerately expanding, it only
 means that, on large scales, the Friedmann-Robertson-Walker
 metric is not the correct metric which describes our universe.

 \section{Conclusions}

 In the first section of this paper, we ask if the
 conventional variable separation techniques in
 the studying of standard cosmology and the collapsing of
 extremely dense stars introduce Newton's absolute space-time
 concepts. In the second section we point out that a maximumly
 symmetric metric may only be used to describe an absolute
 space-time in which the signal transferring speed can be
 infinitely large.
 In our real universe, the speedestly transferring signal is light,
 so its describing metric cannot have the maximumly symmetric properties.

 In the third section of the paper, we build the basic frame-work
 of a completely raltive cosmology, in which the cosmological
 principle is expressed consistently with special relativity. In
 standard cosmology, this is not the case. In the fourth section,
 we provide the basic equations controlling the evolutions of the
 quantities involved in our completely relative cosmology. In the
 fifth section, we prove that in a completely relative cosmology,
 the current observed luminosity-distance v.s. red-shift relations
 of supernovaes may be explained naturally without assuming that
 our universe is accelerately expanding.

 If what we criticized here of standard cosmology is the fact,
 i.e., Friedmann-Robertson-Walker metric cannot be the correct
 metric describing our universe on large scales and long time
 evolution processes, then we will have to change it
 revolutionally. After the changes, we think at least four
 problems in our current cosmology will appear differently or even
 will not appear at all: (i)dark energy and cosmological constant
 problem, (ii)the primordial singularity problem, (iii)horizon
 problem and flattness problem and (iv) the necessity of
 inflations. Probably, the singularity problem of black holes will
 have different appearrance either, because our
 current black hole formation theory is
 based on the extremely dense stars collpasing.

 Alghough we have not get definite solutions from the basic equations we
 provide in the fourth section of this paper, we think such
 solution must exist. While the tasks of searching for such
 solutions must be challengeful and meaningful, both physically
 and mathematcally. It is a completely new area of general
 relativity and cosmology.

 {\bf About references} about observations which indicate that
 our universe is accelerately expanding and an non-zero
 dark energies existing, we referred to
 \cite{Riess98, Perlmutter98, Knop03, Tonry03, Riess04, WMAP03, Tegmark04};
 about theoretical studying of dark energies, we
 referred to \cite{Quintessence1, Quintessence2, Phantom,
 backReaction}; about inflation and Ekpyrotic universes we
 referred to \cite{Guth81, Linde82, EkpyroticUniverse}; about the
 non-factorizable solutions of Eintein equaitons we referred to
 \cite{SWeinberg, Assumption, Sphere}, professor A.D.Linde thinks
 that the solutions we provided in \cite{Assumption} may have
 relevance to \cite{SelfReproduceUniverse}. We thank very much to
 professor A.D.Linde to inform us this point.

\end{document}